# Confinement Characterization of a 2-Electron Quantum Dot

Preston Jones

Physics Department

Virginia Military Institute

Lexington, Virginia 24450

November 24, 2003

**Abstract:** The time independent Schoedinger equation for two electrons confined in a parabolic external potential is solved. Developing this solution in terms of a dimensionless variable it is demonstrated that parameterization of the strength of the confining potential separate from the effective mass assumption greatly clarifies the functional dependence of the system energy on the system parameters. The determination of the strength of the external confinement and validation of the effective mass assumption in real devices is greatly improved by characterizing the strength of the confining potential separate from the effective mass.

## Introduction

Reaching a better understanding of nanoscopic systems in general and quantum dots in particular requires the development of good models for interacting particles in an external confining potential. Toward this end a great deal of recent research has focused on the model system of two interacting electrons confined in a parabolic potential. This model system is sometimes referred to as artificial helium and, as with real helium, the system consist of two interacting electrons in a confining potential. Though somewhat idealized, this system presents an extremely useful model for studying quantum effects in nanoscopic systems. Unlike a real helium atom, artificial helium does not offer a precisely known external confining potential apriori, making the accuracy of the characterization of the external potential highly dependent on the quality of the theoretical model of the system. In the present study it is demonstrated that the ubiquitous practice of modeling the external confining potential of quantum dots by a characteristic frequency makes it impossible to determine the relative contribution of the effective mass to the energy eigenstates independent from the strength of the potential.

The quantum many-body problem of the parabolic confining potential has been useful in both improving the understanding of the physics of quantum dots [Laufer] [Kais] [Taut] [Pfannakuche] [Rontani] [Szafran] as well as characterizing the external confinement of specific devices [Rontani] [Tarucha]. To further improve the understanding of quantum dots the Schrödinger equation of the idealized system of artificial helium is solved and presented in a form that is particularly well suited for evaluation of the parameters that characterize the parabolic confining potential in general or the confinement of electrons in a specific device. This solution also offers an efficient and accurate method for electronic structure calculations and the construction of high quality geminal state functions.

Previous studies of electron confinement in quantum dots have generally followed the practice of characterizing the confining potential by a frequency parameter and assuming the effective mass of the periodic lattice. At some point in the reduction of the dimensions of electron confinement the effective mass must become equal to the mass of a free electron as the system transitions from lattice to cluster to molecule [Gaponenko]. Coupling the characterization of the confining potential together with the effective mass in the frequency parameter makes it extremely difficult or impossible to evaluate this transition as the feature size of the quantum dot is reduced or as the strength of the confining potential is increased.

The quantum many-body problem of two electrons confined in a parabolic potential was previously solved using the basis method by Lamouche and Fishman [Lamouche]. Expanding upon this method of solution the eigenvalues and statefunctions of artificial helium have been developed in a particularly useful form that clearly illuminates the dependence of the energy eigenvalues on the characterization of the external potential as well as the effective mass. In this form the solution is sufficiently general to permit quantitative consideration of various assumptions of the system parameters for model systems and real devices. Development of the solution in this form further permits qualitative as well as quantitative studies of the eigenvalues of this system demonstrating that the characterization of the confining potential is greatly improved by separating the parameter describing the potential from that of the effective mass.

## Two Electrons in a Parabolic External Potential

A system of two interacting electrons confined in a quantum dot is described by the state function which is the solution to the 2-body Schrödinger equation, $H\psi(1,2) = E\psi(1,2)$. The system energies $E$ are the eigenvalues of this equation. The Hamiltonian in absolute coordinates for the model parabolic external confining potential is written as

$$H = -\frac{\hbar^2}{2m_{eff}}\nabla_1^2 + \frac{1}{2}kr_1^2 - \frac{\hbar^2}{2m_{eff}}\nabla_2^2 + \frac{1}{2}kr_2^2 + \frac{e^2}{\kappa|\mathbf{r}_2 - \mathbf{r}_1|}.$$

One of the reasons that this problem has proven to be so useful is that the Schrödinger equation divides into two separate parts with a judicious change of coordinate. In particular the equation in absolute coordinates is transformed to relative (rel), $\mathbf{r} = \mathbf{r}_2 - \mathbf{r}_1$, and center of mass (CM), $\mathbf{R} = \frac{1}{2}(\mathbf{r}_2 + \mathbf{r}_1)$, coordinates. The transformation of the Hamiltonian is accomplished by simple substitution $(\mathbf{r}_1, \mathbf{r}_2) \to (\mathbf{r}, \mathbf{R})$.

After making the substitution for the absolute coordinates the sum of the Laplacians in the Hamiltonian is rewritten in the new coordinates as $\nabla_2^2 + \nabla_1^2 = 2\nabla_\mathbf{r}^2 + \frac{1}{2}\nabla_\mathbf{R}^2$. The parabolic external confining potential requires only a small rearranging of terms and the interaction part becomes particularly simple. Collecting the various pieces in the rel and CM coordinates the Hamiltonian can be rewritten as

$$H = -\frac{\hbar^2}{4m_{eff}}\nabla_\mathbf{R}^2 + kR^2 - \frac{\hbar^2}{m_{eff}}\nabla_\mathbf{r}^2 + \frac{1}{4}kr^2 + \frac{e^2}{\kappa r}.$$

With this change of coordinates the eigenvalue problems in the relative and center of mass coordinates, $H = H_\mathbf{R} + H_\mathbf{r}$, can be solved independent from each other. The state function for the Hamiltonian in these new coordinates is rewritten as the product of separate functions of the two halves of the Hamiltonian and the spin function, $\psi(1,2) = \varphi(\mathbf{r})\xi(\mathbf{R})\sigma(s_1,s_2)$.

The part of the 2-body problem that is a function of the CM coordinate, $H_\mathbf{R}\xi(\mathbf{R}) = \eta\xi(\mathbf{R})$, can be written in a familiar form as

$$\left(-\frac{\hbar^2}{2m_{eff}}\nabla_\mathbf{R}^2 + \frac{1}{2}k_R R^2\right)\xi(\mathbf{R}) = \eta'\xi(\mathbf{R}),$$

where $\eta' = 2\eta$ and $k_R = 4k$. This is the exact form of the eigenvalue equation for the isotropic 3-dimensional linear harmonic oscillator and the solution is well known [Ohanian, Merzbacker]. The eigenvalues of this CM equation are $\eta = \hbar\sqrt{\frac{k}{m_{eff}}}\left(2N + L + \frac{3}{2}\right)$. The equation for $\varphi(\mathbf{r})$ and the second term of the Hamiltonian, $H_\mathbf{r}\varphi(\mathbf{r}) = \varepsilon\varphi(\mathbf{r})$, while complicated by the Coulomb interaction, can be similarly rewritten as

$$\left(-\frac{\hbar^2}{2m_{eff}}\nabla_\mathbf{r}^2 + \frac{1}{2}k_r r^2 + \frac{e^2}{2\kappa r}\right)\varphi(\mathbf{r}) = \varepsilon'\varphi(\mathbf{r}),$$

where $\varepsilon' = \frac{1}{2}\varepsilon$ and $k_r = \frac{1}{4}k$. The eigenvalues of the 2-body Schrödinger equation and the total energies of the system are given by the sum of the two eigenvalues,

$$E = \eta + \varepsilon = \frac{1}{2}\eta' + 2\varepsilon'.$$

### Solving the Center of Mass Equation

The CM equation, $H_\mathbf{R}\xi(\mathbf{R}) = \eta\xi(\mathbf{R})$, is the well known isotropic 3-dimensional harmonic oscillator problem. The eigenfunctions for this equation are a product of a function of the magnitude of $\mathbf{R}$ and the spherical harmonics, $\xi(\mathbf{R}) = \xi(N,L)Y_L^M(\hat{\mathbf{R}})$. Writing the CM eigenfunctions in this form permits the separation of the equation leaving only an equation in terms of the magnitude $R$. The eigenvalue problem remaining for the CM equation is

$$\left(-\frac{\hbar^2}{2m_{eff}}D_R^2 + \frac{1}{2}k_R R^2\right)\xi(N,L) = \eta'\xi(N,L),$$

$$D_R^2 \equiv \frac{1}{R^2}\frac{\partial}{\partial R}R^2\frac{\partial}{\partial R} - \frac{L(L+1)}{R^2}.$$

The normalized eigenfunctions of this equation are available from the literature [Rontani],

$$\xi(N,L) = \lambda^{\frac{3}{4}}\sqrt{\frac{2N!}{\Gamma\left(N+L+\frac{3}{2}\right)}}u^{\frac{L}{2}}e^{-\frac{u}{2}}L_N^{L+\frac{1}{2}}(u),$$

where $\lambda = \frac{2}{\hbar}\sqrt{m_{eff}k}$, $u = \lambda R^2$, $L_N^{L+\frac{1}{2}}$ are the associated Laguerre polynomials, and $\Gamma$ the gamma function.

## Solving the Relative Equation

The relative equation is in the form of a 1-particle time independent radial Schrödinger equation and there are various methods that can be used to solve this eigenvalue problem. In particular the equation can be exactly solved using the Frobenious method for specific values of $n$ and $l$ and associated specific values of the confining potential $k$ [Taut]. To solve the problem in general the eigenfunctions of the rel equation are expanded in a basis, converting the differential equation to a matrix eigenvalue problem. This in principle leads to a matrix equation of infinite dimensions which must be truncated to some finite dimension in order to make the problem tractable. By expanding the eigenfunctions in terms of a basis of the eigenfunctions of the corresponding problem without the interaction term the solution has a particularly useful form. In this form the relationship between the parameters of the model confining potential and the energies of the system are clearly demonstrated.

As with the CM equation the eigenfunctions of the rel equation, $H_r \varphi(\mathbf{r}) = \varepsilon \varphi(\mathbf{r})$, can be separated into the product of a spherical harmonic function and a function of the magnitude of the rel coordinate, $\varphi(\mathbf{r}) = \varphi(n,l,m) = \varphi(n,l) Y_l^m(\hat{\mathbf{r}}) = \varphi(r) Y_l^m(\hat{\mathbf{r}})$. After rewriting the eigenfunctions the radial part of the eigenvalue problem becomes

$$\Delta \varphi(r) = \varepsilon' \varphi(r),$$

$$\Delta \equiv -\frac{\hbar^2}{2m_{eff}} D_r^2 + \frac{1}{2} k_r r^2 + \frac{e^2}{2\kappa r},$$

$$D_r^2 \equiv \frac{1}{r^2} \frac{\partial}{\partial r} r^2 \frac{\partial}{\partial r} - \frac{l(l+1)}{r^2}.$$

The eigenfunctions of this equation are expand as $\varphi(n,l) = \sum_{q=0}^{\infty} c_{nq} \chi(q,l)$ where the basis function are similar to those found for the CM equation with the appropriate change in quantum numbers,

$$\chi(q,l) = \gamma^{\frac{3}{4}} \sqrt{\frac{2q!}{\Gamma\left(q+l+\frac{3}{2}\right)}} v^{\frac{l}{2}} e^{-\frac{v}{2}} L_q^{l+\frac{1}{2}}(v),$$

where $\gamma = \frac{1}{2\hbar} \sqrt{m_{eff} k}$ and $v = \gamma r^2$. With the judicious use of closure, the eigenvalue problem is rewritten as a matrix equation, with infinite dimensions, and the components of the matrix form of the differential operator $\Delta$ are

$$I_{pq} = \langle \chi(p,l) | \Delta | \chi(q,l) \rangle.$$

To improve computational efficiency the integrals in the matrix elements $I_{pq}$ are transformed to the dimensionless variable $v$. This transformation is facilitated by first changing the differential volume element,

$$r^2 dr = \frac{1}{2} \sqrt{\frac{v}{\gamma^3}} dv,$$

and differential operator,

$$\frac{1}{r^2}\frac{\partial}{\partial r}r^2\frac{\partial}{\partial r}f(v) = 6\gamma\frac{\partial}{\partial v}f(v) + 4\gamma v\frac{\partial^2}{\partial v^2}f(v),$$

to the dimensionless variable. The transformed radial differential operator is conveniently expanded using the above identities and recalling that $\gamma = \frac{1}{2\hbar}\sqrt{m_{eff}k}$ and $k_r = \frac{1}{4}k$, yielding

$$\Delta = -\frac{\hbar^2}{2m_{eff}}D_r^2 + \frac{\hbar}{4}\sqrt{\frac{k}{m_{eff}}}v + \frac{1}{2\sqrt{2}}\frac{e^2}{\kappa\sqrt{\hbar}}\sqrt{\sqrt{m_{eff}k}}\frac{1}{\sqrt{v}}.$$

The first term of the operator is expanded as

$$-\frac{\hbar^2}{2m_{eff}}D_r^2 = -\frac{3}{2}\hbar\sqrt{\frac{k}{m_{eff}}}\frac{\partial}{\partial v} - \hbar\sqrt{\frac{k}{m_{eff}}}v\frac{\partial^2}{\partial v^2} + \frac{1}{4}\hbar\sqrt{\frac{k}{m_{eff}}}\frac{l(l+1)}{v}.$$

Now the integrals of each matrix elements can be written explicitly as

$$I_{pq} = \frac{1}{2}\int_0^\infty \sqrt{v}\chi(p,l)\Delta\chi(q,l)dv,$$

where $\chi(q,l) = \gamma^{\frac{3}{4}}\chi'(q,l)$. Each component $I_{pq}$ is the product of various terms involving the parameters of the system and a set of integrals that are the same for all systems. These integrals can be readily solved and since they are functions of a dimensionless variable they need be solved only once.

    The eigenvalue problem is now a matrix problem of infinite dimension. In order to solve the problem numerically the matrix and the expansion of the eigenfunctions must be truncated to some finite number of basis functions $Q$ and then the corresponding finite matrix eigenvalue problem can be solved by standard methods. In matrix form the truncated eigenvalue equation is written as $[I_Q][\varphi_Q(n,l)] = \varepsilon_Q'[\varphi_Q(n,l)]$. The eigenvalues $\varepsilon' \cong \varepsilon_Q'$ are found by diagonalizing the matrix $[I_Q]$ where the elements of the matrix are $I_{pq}$.

## State Functions

Associated with each eigenvalue $\varepsilon_Q'$ of the finite eigenvalue problem in matrix form is a corresponding eigenvector $[\varphi_Q(n,l)]$. The components of these eigenvectors are the coefficients, $c_{nq}$, which approximates the eigenfunctions of the rel equation,

$$\varphi_Q(n,l) = \sum_{q=0}^{Q} c_{nq} \chi(q,l).$$

With this approximation to the eigenfunctions of the rel equation the complete state function is approximated as

$$\psi(1,2) \cong \varphi_Q(n,l) Y_l^m(\hat{\mathbf{r}}) \xi(N,L) Y_L^M(\hat{\mathbf{R}}) \sigma(s_1, s_2).$$

To insure that the statefunctions are normalized each coefficient in the expansion can be divided by the coefficient $c_{nn}$, $c_{nq} \rightarrow \dfrac{c_{nq}}{c_{nn}}$. The solution to the 2-electron problem is not complete without ensuring that the state function is antisymmetric, which is necessary to satisfy the exclusion principle. The construction of an antisymmetric state function for this system of fermions, noting that $Y_l^m(-\hat{\mathbf{r}}) = (-1)^l Y_l^m(\hat{\mathbf{r}})$, leads immediately to a restriction on the angular momentum quantum numbers $l$ and the spin states [Taut]. The complete state function will be antisymmetric for even $l$ only if the spin functions are antisymmetric; $\sigma_A(s_1, s_2) \rightarrow l = 0, 2, ...$, and for odd $l$ only if the spin functions are symmetric; $\sigma_S(s_1, s_2) \rightarrow l = 1, 3, ...$. The antisymmetric spin function can be associated with a singlet state while the symmetric spin function can be associated with the triplet state.

## Characterization of the External Potential

Assuming that the confining potential is known apriori, the characterization of the confining potential by a frequency parameter $\omega$ presents no difficulty. However, determination of the confinement in real quantum dots must be made by the comparison of electronic structure calculations with observations [Tarucha]. In turn the electronic structure calculations are highly dependent on the effective mass. In these calculations it has been widely assumed that the effective mass of the confined electrons is the same as that of electrons in the bulk crystal. Since the effective mass is based on the state functions of the periodic crystal this assumption must break down at some feature size [Gaponenko]. The assumption must also break down for any external confinement where the actual state functions differ greatly from those of the periodic potential of a crystal lattice.

Parameterization of the external potential by a characteristic frequency, $\omega$, can make it very difficult or impossible to determine the validity of any assumptions about the magnitude of the effective mass. This difficulty is due to the weak functional dependence of the ground state energy on variations in $m_{eff}$, independent of $\omega$. In the CM equation the effective mass dependence is completely hidden in the parameter $\omega$ and it is not possible to evaluate this contribution to the ground state energy. For the rel equation inspection of the components of the matrix representation of the differential operator illustrates that the only functional dependence on $m_{eff}$ is in the interaction term and then only to the one quarter power. Substantially all influence of the effective mass is hidden in the characterization of the confining potential by $\omega$.

The functional dependence of the ground state energy on $m_{eff}$ is shown graphically in Figure 1. The ground state energy $E$, with $l = 0$, was first calculated for a constant effective mass and various values of $k$. This plot demonstrates the dependence of the calculated ground state energy on the parameter $k$. The second plot assumes a constant value for $\omega$ and varying $m_{eff}$ demonstrating the very weak functional dependence of the ground state energy on the effective mass separate from the characterization of the confining potential by the parameter ω. The final plot again assumes a constant confining potential parameterized this time by $k$ and exhibits the effect of changes in $m_{eff}$ on the ground state energy calculations.

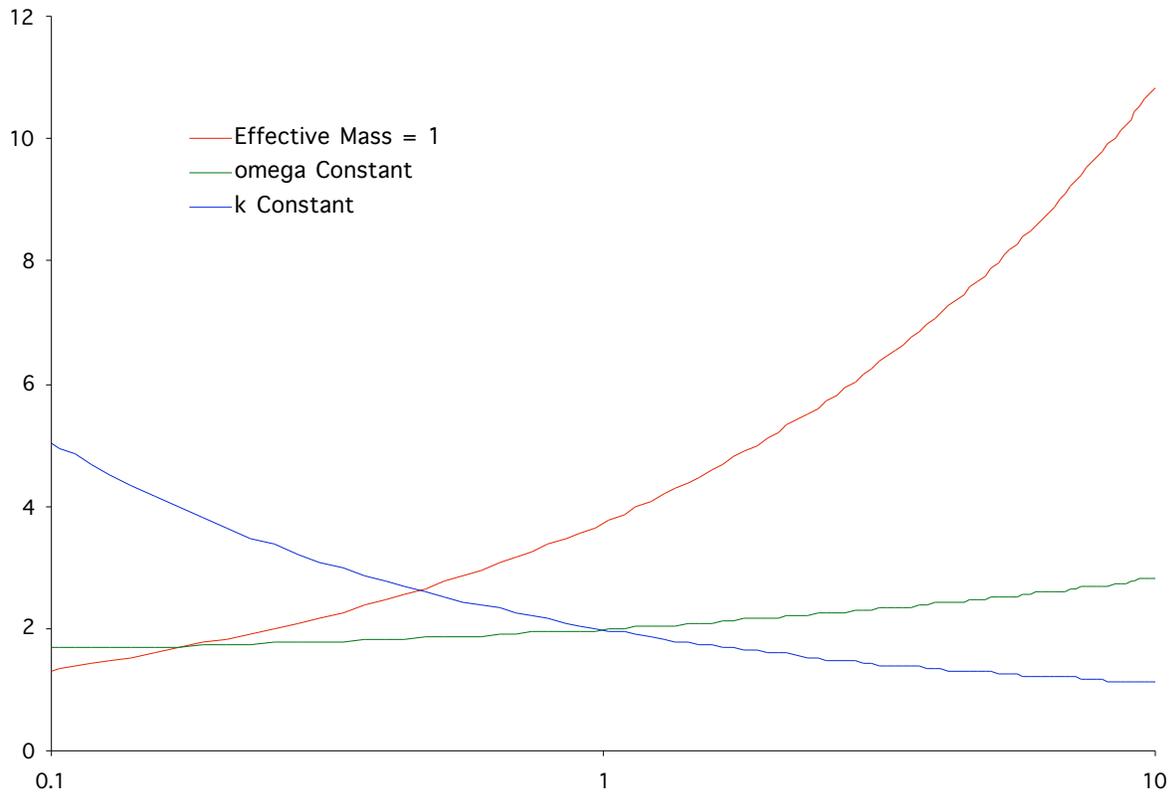

Figure 1. Three plots are shown for $E$ the calculated ground state energy. The "Effective mass = 1" plot is for constant effective mass and varying confining potential characterized by the parameter $k$. The "omega Constant" plot is for constant ω and varying $m_{eff}$. The "$k$ Constant" plot is for constant $k$ and varying $m_{eff}$. All unit are *a.u.*.

Characterization of the external confining potential in terms of the parameter $k$ should make it possible to test the validity of the effective mass for various devises and strengths of external confinement. Figure 2 shows ground state energy $E$ calculations, with $l=0$, for various practical confinement strengths [Tarucha]. The plots for various confinement strengths are compared for the effective mass and the limiting case of the free mass of an electron. In terms of the parameter $k$ the plots are sufficiently distinguishable that it should be possible to test the validity of the effective mass approximation for specific devices and confinement strengths.

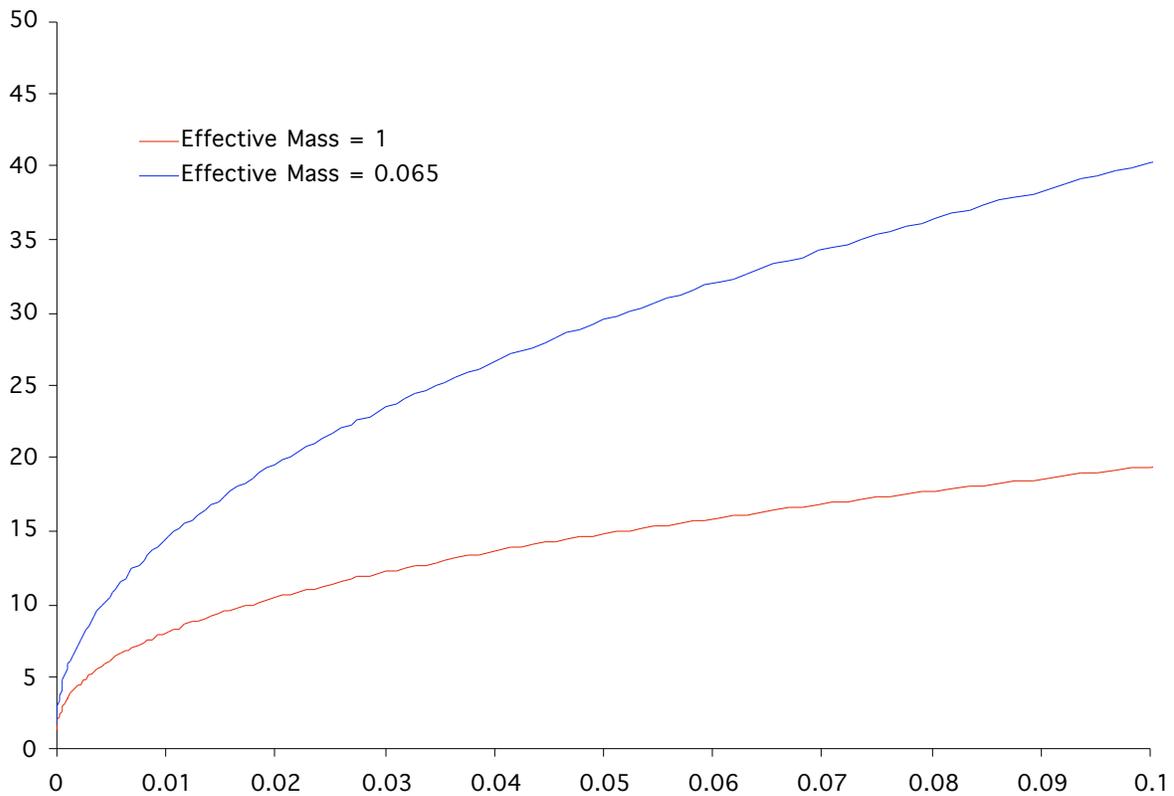

Figure 2.  Two plots are shown for $E$ the calculated ground state energy in $meV$.  Both plots are of constant $m_{eff}$ and varying confining potential characterized by the parameter $k$.   The parameter $k$ is in units of $\frac{meV}{nm^2}$.

## Summary and Conclusions

Recognizing that as feature size and electron confinement in quantum devices become small the effective mass must for some feature size and confining strength in real devices transition to the free electron mass, and that energy calculations for these devices are highly dependent on the assumptions of the model confinement, it has been demonstrated that confinement characterization is greatly improved by parameterizing the confining potential separate from the effective mass.

Toward this end the time independent Schrödinger equation for two electrons confined in a parabolic external potential has been solved using a basis method and presented in an efficient form for computation. This method of solving the rel equation was particularly useful for examining the relationship between the characterization of the strength of the confining potential and the effective mass. It was demonstrated that parameterization of the strength of the confining potential separate from the effective mass clarifies the functional dependence of the system energy on the system parameters.

## References

Pinchus M. Laufer and J. B. Krieger, "Test of density-functional approximations in an exactly soluble model", Physical Review A, 33, pp. 1480-1491 (1986).

S. Kais, R. D. Levine, and D. R. Herschbach, "Dimensional scaling as a symmetry operation", J. Chem. Phys., Vol. 91, No. 12, pp. 7791-7796 (1989).

M. Taut, "Two electrons in a external oscillator potential: Particular analytic solutions of a

Coulomb problem", Physical Review A, Vol. 48, No. 5, pp. 3561-3566 (1993).

Daniela Pfannakuche, Vidar Gudmundsson, Peter A. Maksym, "Comparison of a Hartree, Hartree-Fock, and an exact treatment of quantum-dot helium", Physical Review B, 47, pp. 2244-2250 (1993).

Massimo Rontani, Fausto Rossi, Franca Manghi, Elisa Molinari, "Coulomb-correlation effects in semiconductor quantum dots: The role of dimensionality", Physical Review B, 59, pp. 10165-10175 (1999).

B. Szafran, J. Adamowski, S. Bednarek, "Electron-electron correlation in quantum dots", Physica E, Vol. 5, pp. 185-195 (2000).

S. Tarucha, D. G. Austing, and T. Honda; R. J. van der Hage and L. P Kouwenhoven, "Shell Filling and Spin Effects in a Few Electron Quantum Dot", Physical Review Letters, Vol. 77, No. 17, p. 3613 (1996).

S. V. Gaponenko, "Optical Properties of Semiconductor Nanocrystals", Cambridge University Press (1998).

G. Lamouche and G. Fishman, "Two interacting electrons in a 3-dimensional parabolic quantum dot: a simple solution", J. Phys., Condens. Matter, 10, 7857-7867, (1998).